\documentstyle[prl,aps,epsf]{revtex}

\begin{document}
\draft

\twocolumn[\hsize\textwidth\columnwidth\hsize\csname@twocolumnfalse\endcsname

 \title{Overcharging by macroions: above all, an entropy effect.}

\author{Marcelo Lozada-Cassou and Felipe Jim\'{e}nez-\'{A}ngeles}

\address{Instituto Mexicano del Petr\'{o}leo, Eje Central L\'{a}zaro C\'{a}rdenas
152, Apartado Postal 14-805, 07730 M\'{e}xico, D. F., M\'{e}xico, and
$^{{\rm} }$Departamento de F\'{\i}sica, Universidad Aut\'{o}noma
Metropolitana-Iztapalapa, Apartado Postal 55-534, 09340 M\'{e}xico, D.F.,
M\'{e}xico.}

\date{\today}

\maketitle

\begin{abstract}
Model macroion solutions next to a charged wall show interface \textit{true overcharging}, charge
reversal and inversion, and layering. Macroion layering is present, even if
the wall or the macroparticle are \textit{uncharged} or if the wall and macroions are
like-charged. An effective long-range attractive force between the adsorbed
macroions is implied. The results are obtained through an integral equation
theory and a new extended Poisson-Boltzmann theory, and are in accordance
with experiments on confined macroions and polymer layering.
\end{abstract}


\vskip1pc] \narrowtext
%
The restricted primitive model (RPM) for an electrolyte solution includes
the two main forces in this system: the long range Coulombic and the
short-range repulsive forces. In RPM the ions are taken to be hard spheres
of diameter $a$ and charge {\it ez}$_{i}$ ($e$ is the protonic charge and $%
z_{i}$ is the ionic valence), embedded in a dielectric medium of dielectric
constant {\it $\varepsilon $}. This model has been shown to be in agreement
with Monte Carlo (MC) simulations and experimental results of bulk and
confined electrolyte systems \cite{attard96}.


When a {\it divalent} electrolyte, at high concentration, is next to a
charged wall, the charge of the adsorbed counterions to the wall overcome
that on the wall. This effect produces a second layer of ions, where the
coions outnumber the counterions. These effects are known as {\it charge
reversal} (CR), some times (perhaps improperly) referred to as overcharging,
and {\it charge inversion} (CI), respectively. Although these phenomena have
been reported, both theoretically \cite {lozada82} and by computer simulations \cite {megen80}, since
1980, important implications to protein electrophoresis \cite {lozada99} and medicine \cite {gelbart00}
were later recognized. On the other hand, the long-range attraction between
confined like-charged macroparticles \cite {likecharges} and the adsorption of macroions
onto oppositely charged \cite {gurovitch99}  or like-charged \cite {decher97} surfaces have received much
attention. The understanding of these phenomena have been recognized as
relevant for the colloid science and technology \cite {schmitz93}, the oil industry, and
molecular self-assembly (e.g., DNA encapsulation) and nano-structured films
(e.g., polyelectrolyte layering) \cite {gelbart00,gurovitch99,decher97}.


Here, we extend the hypernetted chain/mean spherical approximation (HNC/MSA)
integral equation \cite {lozada96a} to be applied to model macroion solutions next to a
charged wall. The HNC/MSA has been proved to be in good agreement with Monte
Carlo data for the electrical double layer (EDL) of closely related models
\cite {degreve98,lozada96c}. Because of the larger macroion's size, this theory is expected to
be even more reliable than for the simple electrolyte case \cite {lozada82,lozada96c}. The
macroparticle is taken as a charged, hard sphere of diameter $a_{M}$,
concentration {\it $\rho $}$_{M}$ and valence $z_{M}$, whereas the little
ions are modeled by the RPM. The wall has uniform surface charge density
{\it $\sigma $}$_{0}$. The wall dielectric constant is chosen to be equal to
that of the solvent, to avoid image potentials. The ionic distribution, as a
function of the distance $x$ from the surface of the wall, gives the
structure of the equilibrium EDL, and is expressed in terms of the
concentration profiles, {\it $\rho $}$_{wi}$($x$) = {\it $\rho $}$_{i}$g$%
_{wi}$($x$). {\it $\rho $}$_{\iota} $ is the bulk concentration,
of the ionic species $i$, and g$_{wi}$($x$) is the species $i$
reduced concentration profile (RCP). The HNC/MSA integral
equations for the RCPs are given by \cite {lozada96a} $g_{wi} (x)
\equiv \exp [ - \beta W_{i} (x)] = \exp {\left[ {\ - \beta \left(
{ez_{i} \psi (x) + J_{i} (x)} \right)} \right]}$. $W_{i}(x)$ is
the potential of mean force, i.e., the effective {\it total}
wall-ion interaction potential. W$_{{\rm i}}$(x) has two
contributions: the electrostatic potential part, given by the mean
electrostatic potential,{\it $\psi $(x)}, plus the short range
repulsive potential part, due to the ionic
size, given by $J_{i} (x)$. Both functions are functionals of {\it $\rho $}$%
_{wi}$($x$). {\it $\beta $}=1/(K$_{{\rm B}}$T), where K$_{{\rm B}}$ is the
Boltzmann constant and T is the absolute temperature. The ion-ion and the
macroion-ion direct interaction potentials are given by a hard-sphere
potential plus the Coulombic potential. In the limiting case of $a$ = 0
HNC/MSA reduces to the integral equation form of a new extended
inhomogeneous Poisson-Boltzmann (PB) theory \cite {attard96,gelbart00,schmitz93} for point ions plus
macroions, next to a charged wall. Since macroions are considered at finite
concentration, this approach is an improvement to the classical PB equation
for confined macroions, at infinite dilution \cite {gelbart00,schmitz93}, where only two macroions
are considered: i.e., in our theory macroion-macroion correlations are
included. For $a$=0 and {\it $\rho $}$_{M}$ =0, we recover the integral
equation version of the classical Gouy-Chapman (GC) theory for point-ions
next to a charged wall \cite {schmitz93}. A point-ion model (PIM) for an electrolyte
solution is like the RPM, but $a$=0.


We have solved HNC/MSA for several values of $Z_{M}$, $a_{M}$,
$\rho_{M}$, $\sigma_{0}$ and salt parameters: $z_{ +}
:z_{{\rm -}} $, {\it $\rho $}$_{\iota} $ and $a$. We calculated g$_{wi}$($x$), {\it $%
\psi $}($x$) and the effective charge density, $\sigma (x) = - {%
\int\limits_{x}^{\infty} {\rho _{el} (y)dy}} $ [10,13] . The charge profile
in the solution is given by $\rho _{el} (x) \equiv {\sum\limits_{m = 1}^{3} {%
ez_{m}} } \rho _{m} g_{m} (x)$, where we have omitted the
sub-index $w$, for notation simplicity. $x$=$a$/2 is the distance
of closest approach, to the wall, of the small ions. Hence,
$\sigma _{0} = - {\int\limits_{a /
2}^{\infty} {\rho _{el} (y)dy}} $, by the electroneutrality condition, and ${%
\sigma} ^{\prime}(x) \equiv {\int\limits_{a / 2}^{x} {\rho _{el} (y)dy}} $
is the charge induced by the wall, on the fluid, between the wall and the
distance $x$ to the wall. Hence, $\sigma (x) \left( {\ \equiv {\sigma}
^{\prime}(x) + \sigma _{0}} \right)$ is the effective or net charge (wall
plus fluid) at the distance $x$ away from the wall. $\sigma (x)$ measures
overcharging, CR or CI at the interface. The effective electrostatic force
on an ion is $f_{i}^{e} (x) \equiv - ez_{i} {\frac{{\partial \psi (x)}}{{%
\partial x}}} = {\frac{{4\pi ez_{i}} }{{\varepsilon} }}\sigma (x)$.
Therefore, $\sigma (x)$ is not only a measure of the over (under)-charging
but also of the wall-particle effective electrical force. $F_{i} (x) = - {%
\frac{{\partial W_{i} (x)}}{{\partial x}}} = {\frac{{1}}{{\beta} }}{\frac{{%
\partial \ln {\left[ {g_{i} (x)} \right]}}}{{\partial x}}}$ is the net
effective, many-body, force between the wall and an ion of species $i$.
Hence, $F_{i} (x) = f_{i}^{e} (x) + f_{i}^{s} (x)$, where $f_{i}^{s} (x)
\equiv {\frac{{\partial J_{i} (x)}}{{\partial x}}}$ contains the
non-electrostatic contributions. The larger the ionic size, the larger the
confinement excluded volume (CEV) and the smaller the accessible volume. In
our case, $f_{i}^{s} (x)$ has the ionic excluded volume contributions. Since
both {\it $\psi $(x)} and $J_{i} (x)$ are functionals of $\rho _{el} (y)$
and are in a non-linear equation, the charge and size correlations are, in
general, {\it not independent} \cite {lozada96a,lozada90a} and,
hence, {\it our theory predicts
that overcharging, if present, is related to both the electrical charge and
the ionic size}. It is an elementary statistical mechanics result that the
smaller the accessible volume the smaller the entropy of the system. Thus,
the larger the ionic size the lower the entropy of the system. Charge
correlations of like-charged ions have the effect of also reducing the
system's accessible volume and, thus, of reducing its entropy.


In all our calculations T=298 K, {\it $\varepsilon $}=78.5 and {\it $\rho $}$%
_{M}$ =0.01 ${\rm M}$. In the RPM calculations $a$=4.25 {\AA}. The
4.25 {\AA} ionic size approximately corresponds to that of a
hydrated ion. In Fig. \ref{Fig1}, the macroion has negative
charge, which is opposite to that of the wall and the divalent
salt ion. The macroion RCP shows a very strong adsorption to the
wall. A second layer of macroions is adsorbed, with an
intermediate layer of divalent positive ions, followed by
monovalent negative little ions. A layer of positive divalent
ions, followed by a monovalent negative ions layer, also mediates
a third layer of adsorbed negative macroions. The macroions of the
first layer are surrounded by counterions. Subsequent layers of,
less concentrated, macroions are observed, also mediated by layers
of positive and negative little ions. Considering that the bulk
macroion concentration is 0.01M, the local macroion concentration
at the second peak, $\approx $0.035M, is not negligible. The first
peak is $\approx $20M. Hence, large macroions, next to a highly
charged wall, {\it assemble} next to the wall. In the inset, at a
distance of one macroion radius, a deep minimum is observed,
corresponding to a very strong CR. The maximum located around
15$a$/2 show a CI. For $x$ lower than one macroion radius, {\it
true overcharging} of the wall is present: That is, wall's
divalent coions, are adsorbed to the wall and their charge exceeds
that of the wall plus the adsorbed negative little ions$.$ This
effect has not been reported before and is probably present only
in macroions solutions. The effective wall electrical field, which
is proportional to $\sigma (x)$, is positive before the first
layer of macroions and then negative, before the second layer.
Hence, the electrical force is first attractive and then repulsive
to negative ions. The behavior of the total force on an ion of
species $i$, however, is implicit in the RCP, i.e., a $g_{i} (x)$
above (below) 1 implies that $F_{i} (x)$ is attractive
(repulsive). At $x=a$/2, $\sigma (x)$ is equal to the wall's
charge, whereas for $x \to \infty $, $\sigma (x) \to 0$, as it
should be if electroneutrality is satisfied. Lower wall charge
density or lower macroion's charge, size or concentration produce
lower adsorption. The 2:1 electrolyte solution does not show CR.

In Fig. \ref{Fig2}, the wall and macroion have negative charge.
Here, the positive divalent little ions are counterions of the
wall and macroion. A layer of positive ions, followed by negative
ions is adsorbed to the wall. Then, a strong adsorption of
macroions is observed. In the inset, a CR is followed by a CI. In
Fig. \ref{Fig1}, the macroions are responsible for the first CR,
whereas, in Fig. \ref{Fig2}, they are responsible for the first
CI. The 2:1 electrolyte does show CR. Thus, if no macroions are
present, monovalent counterions do not produce CR, whereas
divalent counterions do. The first
peak corresponds to a 0.2 M macroion concentration. This implies a {\it %
long-range} effective attraction, between the adsorbed macroions. This
attraction is lower, as the macroions-wall distance increases or for smaller
macroions. Our results show that the macroion adsorption depends on $\rho
_{T}^{\ast} \equiv \rho _{ +} a_{ +} ^{3} + \rho _{ -} a_{ -} ^{3} + \rho
_{M} a_{M}^{3} $, i.e., it depends on the CEV. Hence, adsorption of larger
macroions implies longer range and more strong effective macroion-macroion
attraction. These findings are suggestive, in relation to experimentally
reported attraction between like-charged macroions, next to a like-charged
wall, where $\mu $m size confined macroions show the same behavior [6].


In Fig. \ref{Fig3}, {\it uncharged}, large particles, immersed
into a 1:1 RPM electrolyte are next to a negatively charged wall.
A slight CR is observed in the inset, even though that the
counterions are monovalent. EDL studies for 1:1 electrolytes show
that there is no charge or RCP oscillations around 1 \cite
{lozada82}. Hence, in Fig. \ref{Fig3}, the oscillations in the
little ions RCP are due to size correlations related to the
macroparticle adsorption to the wall. Thus, macroion charge is not
necessary to have macroparticle adsorption and their presence
induce ionic oscillations for 1:1 electrolytes.


In Fig. \ref{Fig4} the wall is {\it uncharged}. Since the macroion
has a larger size, $a_{M}$=6.5$a$, we have increased the
macroion's charge to have the same macroion's surface charge
density. The result is a strong adsorption of macroions to the
wall. The local macroion concentration, at contact with the
wall is $\approx $6.63M. In the inset, strong positive and negative {\it true%
} {\it overcharging} of the wall is seen. If $a_{M}$=4.5$a$ or
$Z_{M}$=-40 (not shown) the maximum of the second layer of
macroions decreases from 1.7, in Fig. \ref{Fig4}, to 1.2, but the
RCP's and $\sigma (x)$ curves are qualitatively equal. If no
macroions are present, $\sigma (x) = 0,\forall x > 0$. Since the
wall is uncharged, the strong macroion adsorption and wall
overcharging is due to the large CEV, imposed by the macroions
size. Thus {\it less accessible volume}, which implies {\it less
entropy}, impose {\it more order} at the interface, i.e., more
adsorption to the wall.


In Fig. \ref{Fig5}, we repeat the calculation of Fig. \ref{Fig4}
but, now, the electrolyte
species have zero diameter (PIM) and the macroion has a smaller diameter, $%
a_{M}$=4.5$a$. This case corresponds to an extension of the PB
theory, where only two macroions are considered
\cite{likecharges}.
 In the inset, before one macroion
radius, positive overcharging is present. The maximum is at $x=a_{M}$/2. The
maximum of negative overcharging is at $x \approx $9.6$a$/2. If no macroions
are present, $\sigma (x) = 0,\forall x > 0$. Notice the oscillations in the
point-ion RCP. This oscillatory behavior is in accordance with the {\it exact%
} second moment condition (SMC) of Stillinger and Lovett \cite {stillinger68}.
 However, it
is well known that the RCPs obtained from the PB equation for PIM
electrolytes in bulk, next to a charged wall or around two
macroions (DLVO theory \cite {gelbart00,schmitz93}), do not show
oscillations and the macroion-macroion interaction is purely
repulsive \cite {attard96,schmitz93}. In fact, it is a
mathematical theorem that the PB equation can not predict and
attractive force for two like-charged macroions, between them or
with the wall, at infinite dilution \cite {neu99}. The adsorption
of macroions to the wall implies an effective attraction between
them and with the wall. The difference of our Fig. \ref{Fig5}
result with the classical PB result is the finite concentration of
macroions, which implies a proper consideration of entropy.


In summary, CR, CI and {\it true overcharging} of a wall depend on
electrostatic interactions and on the CEV, which depends on $\rho
_{T}^{\ast} \equiv \rho _{ +} a_{ + }^{3} + \rho _{ -} a_{ -} ^{3} + \rho
_{M} a_{M}^{3} $ (larger $\rho _{T}^{\ast} $ implies more adsorption). On
one hand: a) Higher wall or macroion charge, enhance adsorption; b) Typical
hydrated monovalent counterions (e.g., Na$^{{\rm +} }$, $a \approx $4.25 {\AA%
}) do not show CR (Fig. \ref{Fig1}), whereas typical divalent
counterions do (Fig. \ref{Fig2}). This is due to a greater
efficiency of divalent counterions to store charge \cite
{messina00}. On the other hand: 1) Overcharging can be present for
uncharged walls or macroions, or for like charged wall and
macroions, provided $\rho _{T}^{\ast} $ is sufficiently large; 2)
Point ions can never overcharge a surface \cite
{attard96,lozada99,schmitz93,lozada90a}; 3) In a pure electrolyte
solution, larger monovalent counterions, such as hydrated
Li$^{{\rm +} }$ ($a \approx $7 {\AA}), show CR \cite {gonzales85}.
Therefore, while that CR, CI and {\it true overcharging} result
from the competition between energy vs. entropy, our results show
that entropy has a much more important role than has been
recognized in the past, and give some insight on the
experimentally found long range effective attraction of
like-charged macroions next to a wall \cite {likecharges}, polymer
layering \cite{gurovitch99}{decher97}, and self-assembled
complexes \cite {gelbart00}.


We thank CONACYT (L007E and C086A) and NEGROMEX.

\newpage
\begin{figure}
\epsfxsize=6.5cm\centerline{\epsfbox{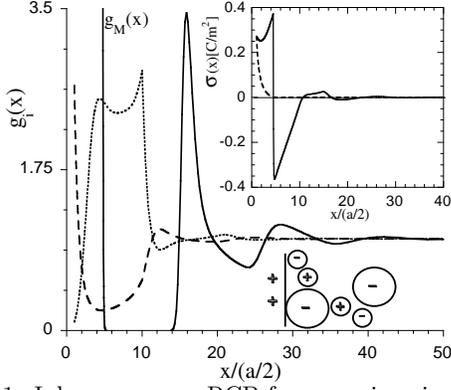}}
\caption{Inhomogeneous RCP for macroions in a 2:1 RPM electrolyte
solution, as function of the distance to the wall $\rho _{M} =
0.01M$, $\rho _{ +}  = 0.7M$, $\rho _{ -}  = 1.0M$, $\sigma _{0} =
0.272C / m^{2}$, $a_{M} = 4.5a$, $Z_{M} = - 40$, $z_{ +}  = 2$,
$z_{ -}  = - 1$. The solid, dash, and dot lines are the macroion
($M$), negative (-) ion, and positive (+) ion RCP, respectively.
In the inset the solid line is the effective charge density
profile, $\sigma (x)$, as a function of the distance to the wall,
for the macroion solution, whereas the dash line is $\sigma (x)$
for a 2:1 RPM electrolyte ($\rho _{ +}  = 0.5M$, $\rho _{ -}  =
1.0M$, $z_{ +}  = 2$, $z_{ -}  = - 1$), when \textit{no} macroions
are present. The sketch roughly reprent the distribution of ions
indicated by their RCP.}
\label{Fig1}
\end{figure}

\begin{figure}
\epsfxsize=6.5cm \centerline{\epsfbox{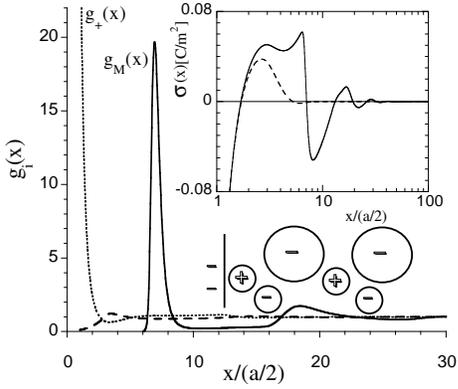}}
\caption{As in Fig. \ref{Fig1}, but , $\sigma _{0} = - 0.272C /
m^{2}$. In the inset, the dash line is $\sigma (x)$ for a 2:1 RPM
electrolyte, when no macroions are present, $\rho _{ +}  = 0.7M$,
$\rho _{ -} = 1.4M$, $z_{ + } = 2$, $z_{ -}  = - 1$.} \label{Fig2}
\end{figure}

\begin{figure}
\epsfxsize=5.5cm \centerline{\epsfbox{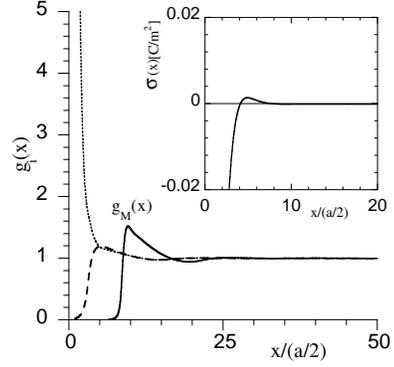}}
\caption{As in Fig. \ref{Fig1} but the salt is a 1:1 RPM
electrolyte solution and $\rho _{ +}  = 1.0M$, $\rho _{ -}  =
1.0M$, $\sigma _{0} = - 0.272C / m^{2}$, $a_{M} = 6.5a$, $Z_{M} =
0$, $z_{ +}  = 1$, $z_{ -}  = - 1$.}\label{Fig3}
\end{figure}

\begin{figure}
\epsfxsize=5.5cm \centerline{\epsfbox{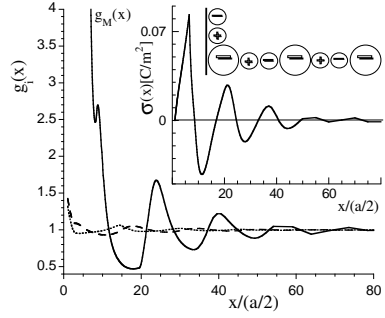}}
\caption{As in Fig. \ref{Fig1} but $\rho _{ +}  = 0.915M$, $\rho
_{ -} = 1.0M$, $\sigma _{0} = 0.0C / m^{2}$, $a_{M} = 6.5a$,
$Z_{M} = - 83$.}\label{Fig4}
\end{figure}

\begin{figure}
\epsfxsize=5.5cm \centerline{\epsfbox{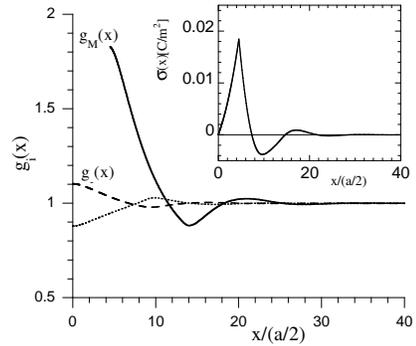}}
\caption{As in Fig. \ref{Fig1} but the salt is a 2:1 PIM
electrolyte solution $\sigma _{0} = 0.0C / m^{2}$, $Z_{M} = -
40$.}\label{Fig5}
\end{figure}

\bibliographystyle{prsty}

\end{document}